\documentstyle[12pt]{article}
\newcommand{\be}{\begin{equation}}
\newcommand{\ee}{\end{equation}}
\newcommand{\bea}{\begin{eqnarray}}
\newcommand{\eea}{\end{eqnarray}}
\newcommand{\nnm}{\nonumber}

\newcommand{\brakket}[3]{\mbox{$\langle #1 | #2 | #3 \rangle$}}
\newcommand{\onehalf}{\mbox{$\scriptstyle 1\over 2$}}
\newcommand{\df}[1]{{\rm d}#1\,}
\newcommand{\dff}[2]{{\rm d}^{#1}#2\,}

\renewcommand{\Re}{\mathop{\rm Re}}
\renewcommand{\Im}{\mathop{\rm Im}}
\newcommand{\mattwo}[1]{\left(\begin{array}{cc}#1\end{array}\right)}
\newcommand{\colvec}[1]{\left(\begin{array}{c}#1\end{array}\right)}
\newcommand{\fvec}{\underline{f}\,}
\newcommand{\ftvec}{\underline{\tilde f}\,}
\newcommand{\ftil}{{\tilde f}}
\newcommand{\ntil}{{\tilde n}}

\newbox\einsbox
\newcommand{\unitmat}{{\setbox\einsbox\vbox{1}
  1\kern-1.5pt\vrule width0.5pt height\ht\einsbox depth0.0pt\relax}}
\newcommand{\dpar}[2]{\frac{\partial#1}{\partial#2}}
\newcommand{\ppar}{p_\parallel}

\title{Pair creation in transport equations
       using the equal-time Wigner function}
\date{January 27, 1993}
\author{
  Christoph Best and J. M.~Eisenberg
  \\[2mm]
  {\small\sl School of Physics and Astronomy,
             Raymond and Beverly Sackler Faculty } \\
  {\small\sl of Exact Sciences, Tel Aviv University, 69978 Tel Aviv,
             Israel,} \\[5pt]
  {\small\sl and} \\[5pt]
  {\small\sl Institut f\"ur Theoretische Physik,
             Johann Wolfgang Goethe-Universit\"at,} \\
  {\small\sl 6000 Frankfurt am Main, Germany} \\
}

\begin{document}
\maketitle
%\tableofcontents

\begin{abstract}
Based on the equal-time Wigner function for the Klein-Gordon field, we
discuss analytically the mechanism of pair creation in a classical
electromagnetic field including back-reaction. It is shown that the
equations of motion for the Wigner function can be reduced to a
variable-frequency oscillator. The pair-creation rate results then from a
calculation analogous to barrier penetration in nonrelativistic quantum
mechanics. The Wigner function allows one to utilize this treatment for the
formulation of an effective transport theory for the back-reaction problem
with a pair-creation source term including Bose enhancement.
\end{abstract}

\section{Introduction}
The usage of the equal-time Wigner transform of the two-point correlation
function of a quantum field theory has been proposed to investigate
nonperturbatively the time evolution of the vacuum state \cite{BGR,Wigner1}.
This approach is based on the Fourier transform of the two-point correlation
function split in the space coordinates but evaluated at equal times. The
equation of motion for this object can be formulated for the interaction of
the Dirac \cite{BGR} or the Klein-Gordon field \cite{Wigner1} with a
classical electromagnetic field and exhibits features of a transport
equation. It does not, in particular, entail a constraint equation---as does
the covariant Wigner function \cite{JME:GK}---and can be computed as an
initial-value problem. This makes this approach a good candidate for
problems like the formation of the quark-gluon plasma and the back reaction
from pair creation in the early universe.

The problem of back-reaction from pair creation in a classical
electromagnetic field has been investigated earlier using the field
equations directly \cite{CoopMott}. Numerical results \cite{Yuval1,Yuval2}
indicate that the exact solution of the field equations can be very well
approximated by a transport theory involving a Schwinger term and Bose
enhancement or Pauli blocking, as pertinent. This has raised the question of
how to derive such a theory consistently starting from the field equations.
While this has been attempted before using the covariant Wigner function
\cite{JME:GK}, we here show how the pair-creation term can be isolated
beginning from the equation of motion for the equal-time Wigner function of
a Klein-Gordon field interacting with a classical electromagnetic field.
Similar approaches based on the equal-time Wigner operator have been taken
in \cite{SG,Yuval3}.

\section{Formalism}
\subsection{The equal-time Wigner function}
The Wigner function of the Klein-Gordon field can be defined from the
symmetrized two-point function
%............................................................................
\be
  C_{\alpha\beta}^+(q,q';t)
  = \brakket{\Omega}{\{\Phi_\alpha(q,t),\Phi_\beta^+(q',t)\}}{\Omega}
\ee
%............................................................................
evaluated at equal times $t$ ($q$ and $q'$ are three-dimensional vectors).
In order to obtain an evolution equation for the Wigner function, it is
necessary to consider the field as a two-component object $\Phi$ in the
Feshbach-Villars representation
%............................................................................
\be
  \Phi = \colvec{\psi \\ \chi}, \qquad
  \phi = \psi + \chi, \qquad
  \left(i\dpar{}{t} - eA^0\right)\phi = m(\psi - \chi),
\ee
%............................................................................
where $\phi$ designates the basic Klein-Gordon field obeying the equation
of motion in a classical electromagnetic field $A_\mu$,
%............................................................................
\be
  \left((\partial_\mu - ie A_\mu)
        (\partial^\mu - ie A^\mu) + m^2\right) \phi(x) = 0.
\ee
%............................................................................
The Wigner function is defined by
%............................................................................
\be
  P(q,p) =
    \int\dff{3}{y} \,
    C^+(q-\onehalf y,q+\onehalf y,t) \,
    \exp\left(\frac{i}{\hbar} p\cdot y + \frac{ie}{\hbar}
              \int_{q-y/2}^{q+y/2} {\cal A}(x)\cdot\df{x}\right),
\ee
%............................................................................
where ${\cal A}(x)$ is the electromagnetic vector potential. The
equation of motion of the Wigner function follows from the equation of
motion for the Feshbach-Villars field
%............................................................................
\be
  i\dpar{\Phi}{t} = \hat H \Phi,
\ee
%............................................................................
with the Hamiltonian
%............................................................................
\be
  \hat H = \frac{(\hat{p} - e{\cal A})^2}{2m} \, a
                  + m \, b
                  + e {\cal A}^0 \unitmat \quad.
\ee
%............................................................................
Here $p$ designates the derivative operator $-i\partial/\partial q$, and
$a$ and $b$ are the matrices
%............................................................................
\be \label{Wig50}
  a = (\sigma_3 + i\sigma_2) = \mattwo{1 & 1 \\ -1 & -1}, \qquad
  b = \sigma_3 = \mattwo{1 & 0 \\ 0 & -1}.
\ee
%............................................................................
In the free case, (\ref{Wig50}) has the solutions
%............................................................................
\be
  \Phi(q,t) = u^{(\pm)}(p) \, e^{\mp i(E_p t - p\cdot q)},
\ee
%............................................................................
with
%............................................................................
\be
  u^{(\pm)}(p) = \frac{1}{2} \sqrt{\frac{m}{E_p}}
                 \colvec{ 1 \pm E_p/m \\ 1 \mp E_p/m },
\ee
%............................................................................
representing particles of positive and negative energy, respectively.

Applying phase-space calculus \cite{Wigner1}, the equation of motion for
$P$ can be written as a differential equation in phase space,
%............................................................................
\bea && \label{Wig49}
  i\left(\dpar{}{t}
         +e{\cal E}(q)\cdot \dpar{}{p} + \dots\right) P(q,p) \nnm\\
  &=&
  i\left[-\frac{p}{2m}\cdot \dpar{}{q}
         +e\left(B(q)\times\frac{p}{m}\right)\cdot\dpar{}{p}
         +\dots\right]
   \left( a\cdot P(q,p) + P(q,p) \cdot a^+ \right) \nnm\\ &&
   {}+\frac{1}{m}\, \left( p^2 -
                    \frac{1}{4} \frac{\partial^2}{\partial q^2} \right)
   \left( a\cdot P(q,p) - P(q,p) \cdot a^+ \right) \nnm\\ &&
   {}+ m\left( b\cdot P(q,p) - P(q,p) \cdot b \right),
\eea
%............................................................................
where the dots indicate quantum corrections involving higher derivatives of
$P$ and of the electric and magnetic field.

The first term on the right-hand side is a (nonrelativistic) flow term. The
second and third terms incorporate relativistic effects (manifested by the
appearance of $p^2$ and $m^2$) and interferences on the Compton scale which
are present in any Wigner function and result from the impossibility to
measure both position and momentum of a quantum-mechanical particle at the
same time. Since the equation of motion for the Wigner function is exact, it
incorporates a mechanism to generate these interferences---namely the term
$\partial^2/\partial q^2$---which can be safely neglected in the
semiclassical limit, i.e., when the spatial variation of the Wigner function
is small on the Compton scale. Note that we do not have to make any
assumption about the variation in time of the Wigner function. This feature
is special to the equal-time Wigner function. This can also be seen from the
fact that, after expanding $P(q,p)$ on a matrix basis (see \cite{Wigner1}
and section \ref{secExp}), $p^2$ appears only in the combination
%............................................................................
\be \label{Wig45}
  p^2+m^2 -
  \frac{1}{4} \frac{\partial^2}{\partial q^2} \approx p^2+m^2 = E_p^2.
\ee
%............................................................................
With this approximation, which is exact for a spatially homogeneous problem,
eq.~(\ref{Wig49}) contains both flow terms as well as local couplings
involving $m$ and $p^2$. In particular, the equations have nonstationary
solutions even in the case of a spatially homogeneous Wigner function. These
oscillations are due to the internal charge degree of freedom of the Wigner
function.

\subsection{Expansion in a diagonalizing basis}
\label{secExp}
To analyze these local oscillations, we expand $P$ into eigenoscillations.
In the free case, two of the eigenoscillations have the frequencies
$\pm 2E_p$, while two others have zero eigenfrequency and correspond to
constant solutions of the equations of motion. The latter can be
interpreted as semiclassical quantities---which later may serve to define
the physical phase-space densities of particles and antiparticles---while
the former represent interferences.

Expanding into eigenvectors of the local oscillations,
%............................................................................
\be
  P(q,p) = \sum_{i=1}^4 f_i \tilde P_i,
\ee
%............................................................................
with the $p$-dependent basis
%............................................................................
\bea \label{Wig47}
  \tilde P_1 &=& \frac{1}{4} \left(\begin{array}{cc}
                 m/E_p + E_p/m    & m/E_p - E_p/m    \\
                 m/E_p - E_p/m    & m/E_p + E_p/m
                 \end{array}\right) \nnm\\
             &=& \frac{1}{2} \left( u^{(+)} \otimes u^{(+)}
                                   +u^{(-)} \otimes u^{(-)} \right), \\
  \tilde P_2 &=& \frac{1}{2} \left(\begin{array}{cc}
                 1                & 0 \\
                 0                & 1
                 \end{array}\right)
              =  \frac{1}{2} \left( u^{(+)} \otimes u^{(+)}
                                   -u^{(-)} \otimes u^{(-)} \right), \\
  \tilde P_{3/4} &=& \frac{1}{4} \left(\begin{array}{cc}
                     m/E_p - E_p/m         & m/E_p + E_p/m \mp 1/2 \\
                     m/E_p - E_p/m \pm 1/2 & m/E_p - E_p/m
                   \end{array}\right) \nnm\\
                 &=&  u^{(\mp)} \otimes u^{(\pm)},
\eea
%............................................................................
the corresponding semiclassical equations of motion for $\fvec$ are
%............................................................................
\bea
  \left(\dpar{}{t} + e{\cal E}(q)\cdot\dpar{}{p}\right) f_1
  &=& -\frac{p}{E_p} \cdot\dpar{}{q} f_2
      + e\frac{{\cal E}(q)\cdot p}{E_p^2} (f_3 + f_4), \\
  \left(\dpar{}{t} + e{\cal E}(q)\cdot\dpar{}{p}\right) f_2
  &=& -\frac{p}{E_p} \cdot\dpar{}{q}
                         \left(f_1+f_3+f_4\right), \\
  \left(\dpar{}{t} + e{\cal E}(q)\cdot\dpar{}{p}\right) f_3
  &=& -\frac{1}{2}\frac{p}{E_p} \cdot\dpar{}{q} f_2 \nnm\\ &&
      {}+ e\frac{{\cal E}(q)\cdot p}{E_p^2} \frac{f_1}{2}
      + 2iE_p f_3, \\
  \left(\dpar{}{t} + e{\cal E}(q)\cdot\dpar{}{p}\right) f_4
  &=& -\frac{1}{2}\frac{p}{E_p} \cdot\dpar{}{q} f_2 \nnm\\ &&
      {}+ e\frac{{\cal E}(q)\cdot p}{E_p^2} \frac{f_1}{2}
      - 2iE_p f_4.
\eea
%............................................................................
Here the approximation of eq.~(\ref{Wig45}) has been made, and the
higher-order quantum corrections to the electromagnetic interaction have
been neglected. The flow term now has the correct relativistic form. The
self-coupling terms that give rise to local oscillations are isolated in
the third and fourth equation (which are complex conjugate to each other),
rendering $f_3$ and $f_4$ oscillatory even in the case of a homogeneous
Wigner function. The expansion into $u^{(\pm)}$ shows that the
nonoscillating terms do not mix positive- and negative-energy solutions,
while the oscillating ones do. The latter oscillations are therefore caused
by Zitterbewegung interferences between particles and antiparticles and
would not be present in a nonrelativistic formulation.

The phase-space densities of observables can be expressed in these
variables:
%............................................................................
\bea
  \mbox{charge:} && e f_2, \label{Wig22} \nnm\\
  \mbox{energy:} && E_p f_1, \nnm\\
  \mbox{current:} && \frac{ep}{E_p} (f_1 + f_3 + f_4), \nnm\\
  \mbox{momentum:} && -p f_2
                    +ip (f_3 - f_4).
\eea
%............................................................................
Note that the current involves the highly-oscillatory components $f_3$ and
$f_4$. It couples to the electromagnetic field and thus provides a mechanism
for the emission of photons in particle annihilation.

A new self-coupling term with the coefficient
%............................................................................
\be \label{Wig48}
  \frac{e{\cal E}(q)\cdot p}{E_p^2}
\ee
%............................................................................
has appeared. This term is due to the fact that the basis (\ref{Wig47})
depends on the coordinate $p$ and therefore generates an additional term
when acted upon by a force term. The physical reason for this term is that
the local oscillations $f_3$ and $f_4$ have different ``directions'' in the
$\fvec$-space at different momenta of the particle. When a particle changes
its momentum because of an external force, its local oscillations will be
slightly out of phase and arouse the $f_1$-component, which can be
interpreted as pair creation. Inspection of (\ref{Wig47}) reveals that the
basis changes most at low momenta, and thus the coupling term is maximal at
$p\ll E$. Physically this means that the pair creation happens at
approximately zero momentum of the particles.

% --- needs work !!
%The equal-time Wigner function is, by definition, not Lorentz-invariant. In
%another reference frame, the Wigner function involves the expectation
%values of the fields at different times. It is not clear that it can be
%constructed from the Wigner function in the original frame of reference
%since phase information may be lost in the transformation. Moreover,
%neither the two-component wave function nor the matrix basis (\ref{Wig47})
%are Lorentz-invariant. This reflects that the definition of particles and
%anti-particles itself is not Lorentz-invariant. On the other hand, since
%the underlying physics is Lorentz-invariant, our description applies to any
%frame of reference which leads to the paradox that pair creation happens at
%approximately zero momentum in {\em any\/} frame of reference. A similar
%paradox is discussed by \cite{Zeldovich}.
% --- needs work

The equal-time Wigner function is, by definition, not Lorentz-invariant. In
another reference frame, the Wigner function involves the expectation values
of the fields at different times. It is not clear that the new function can
be constructed from the Wigner function in the original frame of reference
since phase information may be lost in the equal-time Wigner transformation.
Moreover, since our description is valid in any inertial system, it leads to
the apparent paradox that pair creation happens at approximately zero
momentum in {\em any\/} frame of reference. A similar paradox is discussed
in \cite{Zeldovich}.

The understanding of these equations can be further facilitated by
considering the linear combinations
%............................................................................
\bea \label{Wig32}
  f_+ = \onehalf(f_1 + f_2), &\qquad& f_1 = f_+ + f_-,\nnm\\
  f_- = \onehalf(f_1 - f_2), &\qquad& f_2 = f_+ - f_-,
\eea
%............................................................................
which give the phase-space densities of particles $f_+$ and antiparticles
$f_-$. Rewriting the above equations yields
%............................................................................
\bea
  \left(
    \dpar{}{t} + e{\cal E}(q)\cdot\dpar{}{p}
    + \frac{p}{E_p}\cdot\dpar{}{q}
  \right) f_+ &=&
  -\frac{p}{2E_p} \cdot \dpar{}{q}(f_3 + f_4)
  \nnm\\ && {}+ e \frac{{\cal E}\cdot p}{2E_p^2} (f_3 + f_4),
  \nnm\\
  \left(
    \dpar{}{t} + e{\cal E}(q)\cdot\dpar{}{p}
    - \frac{p}{E_p}\cdot\dpar{}{q}
  \right) f_- &=&
  \frac{p}{2E_p} \cdot \dpar{}{q}(f_3 + f_4)
  \nnm\\ && {}- e \frac{{\cal E}\cdot p}{2E_p^2} (f_3 + f_4),
  \nnm\\
  \left(
    \dpar{}{t} + e{\cal E}(q)\cdot\dpar{}{p}
  \right) f_3 &=&
  -\frac{1}{2} \frac{p}{E_p}\cdot\dpar{}{q} f_2 \nnm\\
  &&{}+ e{\cal E}\cdot\frac{p}{E_p^2} \frac{f_1}{2}
      + 2iE_p f_3, \nnm\\
  \left(
    \dpar{}{t} + e{\cal E}(q)\cdot\dpar{}{p}
  \right) f_4 &=&
  -\frac{1}{2} \frac{p}{E_p}\cdot\dpar{}{q} f_2 \nnm\\
  &&{}+ e{\cal E}\cdot\frac{p}{E_p^2} \frac{f_1}{2}
      - 2iE_p f_4,
\eea
%............................................................................
which exhibits clearly a flow term for $f_+$ and $f_-$ on the left-hand
side. The components $f_3$ and $f_4$ do not have a flow term for position
space since the relevant interferences are not localized. The right-hand
side contains quantum effects and, in particular, pair creation. In the
nonrelativistic limit, $f_+$ and $f_-$ change on a time scale much larger
than the frequency of $f_3$ and $f_4$, so that the right-hand sides average
out and the equations describe a relativistic free gas of positively and
negatively charged particles. Note that $f_-(p)$ refers to antiparticles of
momentum $-p$ which is consistent with the usual hole interpretation.

\section{Pair creation}
\subsection{Reduction to a variable-frequency oscillator}
Let us now investigate the case of a homogeneous constant field, that is,
the Schwinger mechanism of spontaneous pair creation. Because of the spatial
homogeneity, the approximations on local interferences made above in
eq.~(\ref{Wig49}) hold exactly. We shall thus be able to derive exact
equations for the Wigner function in this case.

We first remove the force terms from the equations of motions by separating
out the classical motion using the method of characteristics. Consider a new
function $\ftvec(p_0,t)$ defined by
%............................................................................
\be \label{Wig43}
  \fvec(p,t) = \int\df{p_0} \ftvec(p_0,t) \delta(p-p(p_0,t)),
\ee
%............................................................................
where the function $p(p_0,t)$ describes the classical motion of a particle
with initial momentum $p_0$,
%............................................................................
\be
  p(p_0,t) = p_0 + \int_{-\infty}^{t} \df{t'} e{\cal E}(t).
\ee
%............................................................................
The phase-space evolution of $\fvec(p,t)$ translates into
%............................................................................
\be
  \left(\dpar{}{t} + e{\cal E}\cdot\dpar{}{q}\right) \fvec(p,t)
  = \int \df{p_0} \dpar{\ftvec(p_0,t)}{t} \delta(p-p(p_0,t)).
\ee
%............................................................................
In this way, the momentum derivative is absorbed into the classical
evolution. Here $\ftvec(p_0,t)$ represents the fate of a phase-space
element that started at $p=p_0$ and moves in phase space according to
Newton's law. It may therefore be called a test particle whose inner degrees
of freedom in the Feshbach-Villars sense are governed by
%............................................................................
\bea \label{Wig34}
  \dpar{\ftil_1}{t} &=& \frac{e{\cal E}\cdot p}{E_p^2} (\ftil_3 + \ftil_4),
    \\
  \dpar{\ftil_2}{t} &=& 0, \\
  \dpar{\ftil_3}{t}
  &=& \frac{e{\cal E}\cdot p}{E_p^2} \frac{\ftil_1}{2} + 2iE_p \ftil_3,
    \\
  \dpar{\ftil_4}{t}
  &=& \frac{e{\cal E}\cdot p}{E_p^2} \frac{\ftil_1}{2} - 2iE_p \ftil_4,
                                                            \label{Wig34a}
\eea
%............................................................................
where $p=p(p_0,t)$ is now time-dependent. The set of partial differential
equations (\ref{Wig47}) is thus reduced to a set of ordinary differential
equations. Instead of solving the flow equation in phase space directly, we
have in this way introduced a `test particle', moving in phase space
according to Newton's law and possessing an internal degree of freedom
governed by (\ref{Wig34})-(\ref{Wig34a}). Accordingly, the equations must be
solved for all possible initial values $p_0$. The function $\ftvec(p_0,t)$
thus computed can then be assembled using (\ref{Wig43}) to give the actual
solution of the system of partial differential equations (\ref{Wig47}).

Due to spatial homogeneity, the charge density vanishes. Initially, only
$\ftil_1$ will be nonzero, but by the action of the electric field, the
interference components $\ftil_3$ and $\ftil_4$ are aroused which in turn
couple
back to $\ftil_1$. The coupling
%............................................................................
\be \label{Wig41}
  e\frac{{\cal E}\cdot p(p_0,t)}{E_p^2}
\ee
%............................................................................
vanishes as $p\to\infty$. At large times, $\ftil_3$ and $\ftil_4$ will
therefore settle down into steady oscillations of increasing frequency
and decouple from $\ftil_1$.

The set of ordinary differential equations (\ref{Wig34}) is equivalent to a
single differential equation in a new variable $\xi(t)$,
%............................................................................
\be \label{Wig35}
  \ddot\xi + E_p^2(t) \xi = 0,
\ee
%............................................................................
through the substitution
%............................................................................
\bea \label{Wig40}
  \ftil_1 &=& \frac{E_p}{2E_p^{(0)}}
             \left( |\xi|^2 + \frac{1}{E_p^2} |\dot\xi|^2 \right), \\
  \ftil_3 + \ftil_4 &=& \frac{E_p}{2 E_p^{(0)}}
             \left( |\xi|^2 - \frac{1}{E_p^2} |\dot\xi|^2 \right), \\
  i(\ftil_3 - \ftil_4) &=& \frac{1}{2 E_p^{(0)}}
             (\xi^*\dot\xi + \dot\xi^*\xi),                 \label{Wig40a}
\eea
%............................................................................
where dots denote derivatives with respect to $t$, and $E_p^{(0)}$ is the
test-particle energy at some fixed reference time.
%The current density is therefore
%%............................................................................
%\be \label{Wig38}
%  j = \frac{pe}{E_p} (\ftil_1 + \ftil_3 + \ftil_4)
%    = \frac{pe}{E_p} \, \frac{E_p}{E_p^{(0)}} |\xi|^2.
%\ee
%%............................................................................
The relationship between pair creation in a homogeneous field and
oscillators of time-dependent frequency has been noticed much earlier from
the wave-function approach to pair creation \cite{MP1,MP2,MP3,MP4}. By using
the solutions given in \cite{MP3}, we will be able to give an
approximate asymptotic solution for the Wigner function.

\subsection{Approximate solution}
The general solution of the variable frequency oscillator (\ref{Wig35}) has
the form
%............................................................................
\be \label{Wig39}
  \xi(t) = \frac{1}{\sqrt{E_p(t)}} C_1 \, e^{-i\tau}
          +\frac{1}{\sqrt{E_p(t)}} C_2 \, e^{i\tau},
\ee
%............................................................................
where $C_1$ and $C_2$ are slowly-varying functions, and $\tau(t)$ is
defined by
%............................................................................
\be \label{Wig37}
  \tau(t) = \int_{t_1}^t \df{t'} E_p(t'),
\ee
%............................................................................
with an arbitrarily chosen reference point $t_1$. Making the Liouville
substitution
%............................................................................
\be \label{Wig51}
  \xi(t) = \frac{1}{\sqrt{E_p}} \chi\left(\tau(t)\right),
\ee
%............................................................................
one obtains the oscillator equation in $\chi(\tau)$
%............................................................................
\be
  \chi'' + \left(1+h(\tau)\right) \chi = 0,
\ee
%............................................................................
where primes denote derivatives with respect to $\tau$, and the function
$h(\tau)$ is
%............................................................................
\be
  h(\tau) = \frac{3}{4} \left(\frac{\dot E_p}{ E_p^2}\right)^2
           -\frac{1}{2} \frac{\ddot E_p}{E_p^3}.
\ee
%............................................................................
Equation (\ref{Wig51}) is the time-dependent Schr\"odinger equation of a
particle in a potential $h(\tau)$. The shape of the potential is determined
by $E_p(\tau)$ and thus by the external forces. In particular, $h(\tau)$
vanishes for times $\tau$ when $E_p(\tau)$ is constant, i.e., when the
external field is switched off. Even in a constant electric field,
$E_p\sim\tau$ for $t\to\pm\infty$, $\tau\sim t^2$ and thus
$h(\tau)\sim\tau^{-2}$. The potential is therefore localized, and the
problem is equivalent to a barrier-penetration problem in quantum mechanics.
Note that for a constant electric field $\cal E$, the barrier potential is
%............................................................................
\be
  h(\tau) = \frac{3}{4} \left(\frac{e{\cal E}\cdot p}{E_p^2}\right)^2,
\ee
%............................................................................
in which the quantity (\ref{Wig48}) resurfaces.

The coefficients $C_1$ and $C_2$ represent the incoming and the reflected
waves, respectively. They are asymptotically constant, and the process is
characterized by a barrier penetration coefficient (in addition to a phase
shift)
%............................................................................
\be
  \rho_0 = \left|\frac{C_2}{C_1}\right|^2, \qquad
  t\to\pm\infty,
\ee
%............................................................................
the index $0$ indicating a transition from the vacuum value. Since this
problem is similar to the problem of barrier penetration in ordinary quantum
mechanics, the calculation of this parameter is accomplished by well-known
methods. The quantity $\rho_0$ will depend on the initial momentum and on
the electric field.

Substituting eq.~(\ref{Wig39}) and $\tau=E_p t$ into the relations
(\ref{Wig40}-\ref{Wig40a}), we obtain
%............................................................................
\bea
  \ftil_1 &=& |C_1|^2 + |C_2|^2 = |C_1|^2 (1+\rho_0), \\
  \ftil_3 &=& 2 C_1^* C_2 \, e^{2iE_p t}, \\
  \ftil_4 &=& 2 C_2^* C_1 \, e^{-2iE_p t}.
\eea
%............................................................................
For the vacuum, we choose
%............................................................................
\be
  C_1 = 1, \qquad C_2 = 0.
\ee
%............................................................................
The relationship between the quantity $\rho$ and the number of particles can
be obtained by considering that, due to the reality of the oscillator
equation, the form  $\xi^*\dot\xi - \dot\xi\xi^*$ is a constant,
%............................................................................
\be
  \xi^*\dot\xi - \dot\xi\xi^* = -2iE_p ( |C_1|^2 + |C_2|^2 )
  = {\rm const} = -2iE_p^{(0)},
\ee
%............................................................................
where at $t\to-\infty$, $C_1=1$ and $C_2=0$ have been chosen. From $\rho =
|C_2/C_1|^2$, it now follows that
%............................................................................
\be
  |C_1|^2 = \frac{1}{1-\rho} \frac{E_p^{(0)}}{E_p(t)},
  \qquad
  |C_2|^2 = \rho \frac{1}{1-\rho} \frac{E_p^{(0)}}{E_p(t)}.
\ee
%............................................................................
In particular, we get
%............................................................................
\be
  |C_1|^2 + |C_2|^2
  = \frac{E_p^{(0)}}{E_p(t)} \frac{1+\rho}{1-\rho},
\ee
%............................................................................
which is related to the resulting current density. Substituting
(\ref{Wig39}) into (\ref{Wig40}), it follows that
%............................................................................
\bea \label{Wig52}
  \ftil_1 &=& \frac{E_p}{E_p^{(0)}}
               \left( |C_1|^2 + |C_2|^2 \right)
           = \frac{1+\rho}{1-\rho}, \\
  \ftil_3 + \ftil_4 &=& \frac{E_p}{2E_p^{(0)}}
      \left( C_1^* C_2 \,e^{-2i\tau} + \mbox{c.~c.} \right), \\
  i(\ftil_3 - \ftil_4) &=&
      \frac{E_p}{4 E_p^{(0)}}
      \left( |C_1|^2 + |C_2|^2 - C_1^* C_2 \,e^{-2i\tau} - \mbox{c.~c.}
      \right).                                               \label{Wig52a}
\eea
%............................................................................
As was to be expected, $\ftil_1$ is asymptotically constant, while the
other two components oscillate because of interferences. The current density
in phase space is
%............................................................................
\be
  j(q,p) = \frac{ep}{E_p} (\ftil_1 + \ftil_3 + \ftil_4).
\ee
%............................................................................
The components $\ftil_3$ and $\ftil_4$ will give rise to fast oscillations.
Since
the current couples to the electromagnetic field, this may cause the
emission of photons (or, in the semiclassical approach, electromagnetic
waves of frequency $2E_p$). For our purposes with coupling to a classical
electromagnetic field, the oscillations tend to cancel out upon measurement,
and the main contribution to the current density in phase space comes from
%............................................................................
\be
  j(q,p) \approx \frac{ep}{E_p} \ftil_1
    = \frac{ep}{E_p} \left(1 + \frac{2\rho}{1-\rho}\right).
\ee
%............................................................................
The vacuum value of this quantity is $ep/E_p$; it represents the charge of
the Dirac sea and must be subtracted to yield a measurable quantity. Thus
the phase-space density of pairs produced is given by
%............................................................................
\be
  n = \frac{\rho}{1-\rho},
\ee
%............................................................................
in full correspondence with eq.~(3.7) of \cite{MP3}.

This yields the following picture of pair creation: Let us assume a constant
field ${\cal E}$ switched on at some time $t_0$. For each initial momentum
$p_0$, we have at $t\to-\infty$ $C_1=0$, $C_2=0$, and thus $\ftil_1=1$,
$\ftil_3=\ftil_4=0$. When the field is switched on, the particle is
accelerated or decelerated. In the latter case it reaches its minimum
velocity at some time, corresponding to the barrier in the equivalent
quantum mechanical problem. In this moment, $C_1$ and $C_2$ change so that
$\ftil_1$, $\ftil_3$, and $\ftil_4$ acquire the values
(\ref{Wig52})-(\ref{Wig52a}). This transition takes place when $p(p_0,t)$ is
minimal, i.e., when its parallel part $\ppar$ vanishes. Looking at the
phase-space distribution $\fvec(p,t)$ we therefore see the characteristic
trough found in \cite{BGR} and which signals the impending pair production
at a constant rate per unit space and time: The phase-space density changes
at $\ppar=0$ which creates the left end of the trough; the change is then
transported away by the electric field, and the edge that appeared when the
field was switched on constitutes the right end of the trough.

When the field is not constant in time, it may happen that a test particle
reaches $\ppar=0$ more than once. We then can observe the effects of Bose
enhancement (or Pauli blocking for fermions). In a very strong electric
field with appreciable pair creation we can assume that the transition takes
place during a very small time. This is comparable to solving the barrier
penetration problem by gluing together plane-wave segments at the barrier
with different amplitudes and phase shifts. The time evolution of $\xi(t)$
can then be thought of as being described by (\ref{Wig39}) with the
coefficients $C_1(t)$ and $C_2(t)$ changing their value abruptly whenever
$p(t)$ is small. Let us assume that any one of these transitions can be
described by
%............................................................................
\be
  e^{-iE_p t} \to
  A \, e^{-i(E_p t + \alpha)} + B \, e^{i(E_p t + \beta)},
\ee
%............................................................................
where the amplitudes $A$, $B$ and the phases $\alpha$, $\beta$ depend on the
field strength at the transition. If a general initial state is given by
%............................................................................
\be
  e^{-iE_p t} + \sqrt{\rho} \, e^{i(E_p t + \delta)},
\ee
%............................................................................
with some positive-state admixture characterized by $\rho$ and a phase
shift $\delta$, this initial state will go into a new state
%............................................................................
\bea
  && \left(A \,e^{-i\alpha} + \sqrt{\rho_0} B \,e^{-i(\beta-\delta)} \right)
    e^{-iE_p t} \nnm\\
  &+& \left(B \,e^{i\beta} + \sqrt{\rho_0} A \,e^{i(\alpha+\delta)} \right)
    e^{iE_p t}.
\eea
%............................................................................
The new positive-state admixture $\rho'$, defined as the ratio of the
square of the amplitudes in front of the oscillatory terms, is accordingly
given by
%............................................................................
\be
  \rho' = \frac{\rho + \rho_0 + 2 \sqrt{\rho \rho_0}
                \cos(\alpha+\delta-\beta)}
               {1 + \rho_0\rho + 2 \sqrt{\rho \rho_0}
                \cos(\alpha+\delta-\beta)}
  \approx \rho + \rho_0 + 2 \sqrt{\rho_0 \rho} \cos(\alpha+\delta-\beta),
\ee
%............................................................................
where $\rho_0=B^2/A^2$ is a characteristic quantity of the transition and
will depend on the electric field strength at the time of the transition.
In a nearly constant field, $\rho$ will be given by the Schwinger term.
As we see, $\rho$ behaves as an additive quantity in the transitions, up to
an interference term. The latter can be assumed to average out when
different phase-space trajectories are considered.

Because of this, we can assume that the time evolution of the quantity
$\rho(t)$ is given by
%............................................................................
\be \label{Wig44}
  \rho(t) = \int_{-\infty}^t \df{t'} \delta(p(p_0,t')) \, \rho_0(t'),
\ee
%............................................................................
where $p(p_0,t')$ is the momentum of the test particle ($\rho(t)$ depends of
course on the parameter $p_0$), and $\rho_0(t')$ is determined by the
electric field at the time $t'$. As a result of this assumption, $\rho(t)$
will be a sum of all contributions coming from the individual transitions
which occur whenever $p(p_0,t')$ is small. The approximation here lies in
neglecting the actual transition process in which $\rho$ changes to
$\rho+\rho_0$, since the duration of the transition can be assumed to be
quite small in strong fields.

\subsection{Constructing a flow equation with source term}
We have thus derived an expression for the phase-space density of created
pairs. Remembering that $\rho=\rho(p_0,t)$ we now consider the
phase-space function
%............................................................................
\be
  \ntil(p_0,t) = \frac{\rho(p_0,t)}{1-\rho(p_0,t)},
\ee
%............................................................................
where the tilde reminds us that the classical motion of the test particle
has been separated out. To transform back to a true phase-space
equation, we take the time-derivative of $\ntil(p_0,t)$,
%............................................................................
\be
  \dpar{\ntil(p_0,t)}{t} = \frac{1}{(1-\rho(t))^2} \dpar{\rho}{t}
  = (1+\ntil(p_0,t))^2 \, \dpar{\rho}{t}.
\ee
%............................................................................
Substituting the time evolution of $\rho(t)$, eq.~(\ref{Wig44}), this is
%............................................................................
\be
  \dpar{\ntil(p_0,t)}{t} = |e{\cal E}(t)| \,
                           \delta\left(p(p_0,t)\right) \rho_0(t).
\ee
%............................................................................
Defining the true phase-space function $n(p,t)$ as in (\ref{Wig43}),
%............................................................................
\be
  n(p,t) = \int\df{p_0} \ntil(p_0,t) \,\delta\left(p-p(p_0,t)\right),
\ee
%............................................................................
the phase-space equation of motion is
%............................................................................
\bea \label{Wig53}
  \left(\dpar{}{t} + e{\cal E}\cdot\dpar{}{q}\right) n(p,t)
  &=& (1+n(p,t))^2 \, |e{\cal E}(t)| \nnm\\ &&\quad\times
    \int\df{p_0} \rho_0(t) \,\delta\left(p(p_0,t\right)
                           \,\delta\left(p-p(p_0,t)\right) \nnm\\
  &=& (1+n(p,t))^2 \, \rho_0(t) \,\delta(p).
\eea
%............................................................................
This is a phase-space flow equation with a Bose-enhanced production term on
the right-hand side. The quantity $\rho_0(t)$, characterizing the
pair-production rate, can be computed either perturbatively (pair production
by oscillating fields) or nonperturbatively (Schwinger term), or by some
combination, yielding an enhanced estimate of the pair-production rate.

It has been found earlier \cite{Yuval1} in the study of the
back-reaction problem that pair creation in a spatially homogeneous
problem can be surprisingly well described by the source term
%............................................................................
\bea \label{Wig54} &&
  (1+2 n(p,t)) \frac{|e{\cal E}|}{2\pi}
               \ln\left[ 1 + \exp\left(-\frac{\pi m^2}{|e{\cal E}|}\right)
               \right] \, \delta(p) \nnm\\
  &\approx&
  (1+2 n(p,t)) \frac{|e{\cal E}|}{2\pi}
               \exp\left(-\frac{\pi m^2}{|e{\cal E}|}\right)
               \, \delta(p) \quad,
\eea
%............................................................................
the difference in the Bose enhancement factors in Eqs.~(\ref{Wig53}) and
(\ref{Wig54}) stemming from the lack of a mechanism in the former that
annihilates pairs in favor of an augmentation of the electric field.
%Our result differs from this by a term proportional to $n(p,t)^2$. Such a
%term corresponds to collisions between particles, and it can be argued that
%it is removed when pair annihilation is taken into account. This process is
%connected to the emission of photons which can be taken into account only
%if the electromagnetic field is treated dynamically. As we have seen above,
%the current contains highly oscillatory components due to interferences
%which lead to emission of photons. The correct pair creation rate can thus
%be expected only if the dynamics of the electromagnetic field are included.
%Since this is not the case in our formalism, we overcount the pair
%production rate.

\section{Summary}
We started from the equation of motion for the Wigner function. This
equation of motion is a direct consequence of the field equations. We have
then derived an approximate solution in terms of a transport equation
involving a Schwinger pair creation and a Bose enhancement term. The Wigner
function exhibits an internal structure causing local oscillations, i.e.,
nonstationary solutions in a spatially homogeneous situation. The form of
these oscillations, corresponding to the internal charge degree of freedom
of the Klein-Gordon field, are different at different momenta of the
particles. Thus in the presence of an external force the oscillations mix
with the physically observable phase-space densities of particles and
antiparticles and give rise to particle creation. This process is described
analogously to a barrier-penetration problem in quantum mechanics. The
semiclassical approximation enables us to calculate the pair creation rate
and to give an estimate for repeated pair creation involving Bose
enhancement.

While it was known earlier that pair creation in an external field can be
described by a variable-frequency oscillator from considering the wave
functions, it is the advantage of the Wigner function that it allows one to
cast this in a phase-space formulation. The use of the equal-time Wigner
function is here especially advantageous over the split-time Wigner function
since it allows us to exploit the time-dependent problem of the
variable-frequency oscillator and thus the known solutions to it in a very
natural way.

Some open questions remain: For spatially inhomogeneous fields, we have only
shown how the homogeneous-field limit can be recovered. By making a
systematic expansion in the field gradient, it should be possible to derive
corrections to this result. We have also restricted ourselves to classical
electromagnetic fields. In a next step, the electromagnetic field should be
treated dynamically. In this approximation, emission of photons, pair
annihilation into photons, and particle interaction enter. Finally, even in
the semiclassical limit, the theory is, in general, in need of
renormalization
\cite{BGR}.

In its current form, the equal-time Wigner function is especially suitable
for the explicit solution of semiclassical problems like particle creation
in definite boundary conditions. The fact that it is not relativistically
invariant makes it more difficult to derive generic features of quantum
field theories from it. Still, a possible extension could be to consider
general $N$-particle Wigner functions constructed from $2N$ field operators.
The resulting hierarchy of equations can then be systematically studied.
\\[10pt]
\noindent{\em Acknowledgements: }
We would like to thank Stefan Graf and Yuval Kluger for fruitful
discussions, and I.~Bialynicki-Birula for pointing out some errors.
C.B. wishes to thank G.~Soff, A.~Sch\"afer, and W.~Greiner.
Support from the German Israel Foundation, the Humboldt-Stiftung, and the
Ne'eman Chair in Theoretical Nuclear Physics at Tel Aviv University is
gratefully acknowledged.

\appendix
\section{Appendix: The Schwinger term, obtained by means of the imaginary time
            method}
Since (\ref{Wig35}) corresponds to the time-independent Schr\"odinger
equation describing tunneling through a potential barrier, $\rho$ is the
reflection coefficient for this problem and can be computed by a standard
method, namely that of imaginary times \cite{MP3,MP4}. In this case the
solution is determined by the singularities of $h(\tau)$ in the complex
plane: If $h(\tau)$ has its singularity at $\tau_0$, the value of the
barrier reflection coefficient is
%............................................................................
\be
  \rho = 4 \cos^2(\pi/2\alpha)\, \exp(-4 \Im \tau_0),
\ee
%............................................................................
where $\alpha$ is deduced from the behavior of $E_p(t)$ near the
singularity,
%............................................................................
\be
  \omega(t) \approx \omega_0 (t-t_0)^{\alpha-1}.
\ee
%............................................................................
In the case of a constant field,
%............................................................................
\be
  E_p(t) = \sqrt{ (p + e{\cal E}t)^2 + m^2 },
\ee
%............................................................................
and the singularities are at $E_p=0$, corresponding to
%............................................................................
\be
  t_0 = \frac{-\ppar \pm im_\perp}{e|{\cal E}|},
\ee
%............................................................................
yielding $\alpha=3/2$. To compute the corresponding value of $\tau_0$, we
choose the integration contour in (\ref{Wig37}) along the real axis from
$t_1$ to $\Re t_0$ and then parallel to the imaginary axis to $\Im t_0$. No
contribution to $\Im\tau_0$ arises along the real axis, so the integral to
do remains
%............................................................................
\be
  \Im \tau_0 = \Im \left.i\int_0^{\Im t_0} \df{(\Im t)}
               E_p(t) \right|_{\Re t=\Re t_0}.
\ee
%............................................................................
Writing $x=\Im t_0$, this is
%............................................................................
\bea
  \Im \tau_0 &=& \Re \int_{0}^{\Im t_0} \df{x} ie|{\cal E}|\,
  \nnm\\ && \quad{}\times
  \left[ x^2
         + 2i\left(\Re t_0 + \frac{\ppar}{e|{\cal E}|}\right)
         - \frac{m_\perp^2}{e^2|{\cal E}|^2}
  \right]^{1/2} \nnm\\
  &=& \mp \pi \frac{m_\perp^2}{4e|{\cal E}|}.
\eea
%............................................................................
The barrier-reflection coefficient is therefore
%............................................................................
\be
  \rho = \exp\left( -\pi \frac{m_\perp^2}{e|{\cal E}|} \right),
\ee
%............................................................................
which agrees with the lowest-order term of the Schwinger formula.

\end{document}